\documentclass[sigconf, screen]{acmart}  %
\AtBeginDocument{%
  }

\copyrightyear{2025}
\acmYear{2025}
\setcopyright{cc}
\setcctype{by}
\acmConference[SC Workshops '25]{Workshops of the International Conference for High Performance Computing, Networking, Storage and Analysis}{November 16--21, 2025}{St Louis, MO, USA}
\acmBooktitle{Workshops of the International Conference for High Performance Computing, Networking, Storage and Analysis (SC Workshops '25), November 16--21, 2025, St Louis, MO, USA}
\acmDOI{10.1145/3731599.3767562}
\acmISBN{979-8-4007-1871-7/2025/11}

\usepackage{tikz}  %
\usepackage{todonotes}
\usepackage{xcolor}
\usepackage{url}
\usepackage{hyperref}
\usepackage{csquotes}
\usepackage{censor}
\usepackage[frozencache,cachedir=.]{minted}
\usepackage{siunitx}

\usepackage{subcaption}
 
\newcommand{\parag}[1]{\vspace{2mm}\noindent\textbf{#1.}\hspace{2mm}}

\begin{document}

\title{Optimizing Microgrid Composition for Sustainable Data Centers}

\author{Julius Irion}
\orcid{0009-0003-9639-7347}
\affiliation{%
  \institution{TU Berlin}
  \city{Berlin}
  \country{Germany}
}
\author{Philipp Wiesner}
\orcid{0000-0001-5352-7525}
\affiliation{%
  \institution{TU Berlin}
  \city{Berlin}
  \country{Germany}
}
\author{Jonathan Bader}
\orcid{0000-0003-0391-728X}
\affiliation{%
  \institution{TU Berlin}
  \city{Berlin}
  \country{Germany}
}
\author{Odej Kao}
\orcid{0000-0001-6454-6799}
\affiliation{%
  \institution{TU Berlin}
  \city{Berlin}
  \country{Germany}
}

\renewcommand{\shortauthors}{Irion et al.}

\begin{abstract}
As computing energy demand continues to grow and electrical grid infrastructure struggles to keep pace, an increasing number of data centers are being planned with colocated microgrids that integrate on-site renewable generation and energy storage. 
However, while existing research has examined the tradeoffs between operational and embodied carbon emissions in the context of renewable energy certificates, there is a lack of tools to assess how the sizing and composition of microgrid components affects long-term sustainability and power reliability.

In this paper, we present a novel optimization framework that extends the computing and energy system co-simulator Vessim with detailed renewable energy generation models from the National Renewable Energy Laboratory's (NREL) System Advisor Model (SAM). 
Our framework simulates the interaction between computing workloads, on-site renewable production, and energy storage, capturing both operational and embodied emissions. 
We use a multi-horizon black-box optimization to explore efficient microgrid compositions and enable operators to make more informed decisions when planning energy systems for data centers.
\end{abstract}



\maketitle

\section{Introduction}
\label{cha:introduction}

The rapid expansion of digital services and the increase of large-scale AI deployments have fueled an unprecedented demand for data center capacity~\cite{Bashir2024Climate, wu2024scalingAIsustainably}. 
Between 2017 and 2021, electricity consumption by Meta, Amazon, Microsoft, and Google more than doubled~\cite{epri2024powering} and the International Energy Agency projects that global data center electricity usage will more than double again by 2030~\cite{iea2025energyai}.
Due to this, operators are now facing critical bottlenecks: Recently, Microsoft and Amazon have put data center expansion plans on hold due to insufficient access to infrastructure and reliable grid power~\cite{telegraph2025microsoft, allsup2025microsoft, reuters2025amazon}. 
Without coordinated efforts to expand grid capacity and improve energy efficiency, future digital services may be limited not by technological progress, but by the availability of power~\cite{lin2024exploding_ai_power}.

\begin{figure}
  \centering
  \vspace{5mm}
  \includegraphics[width=.95\columnwidth]{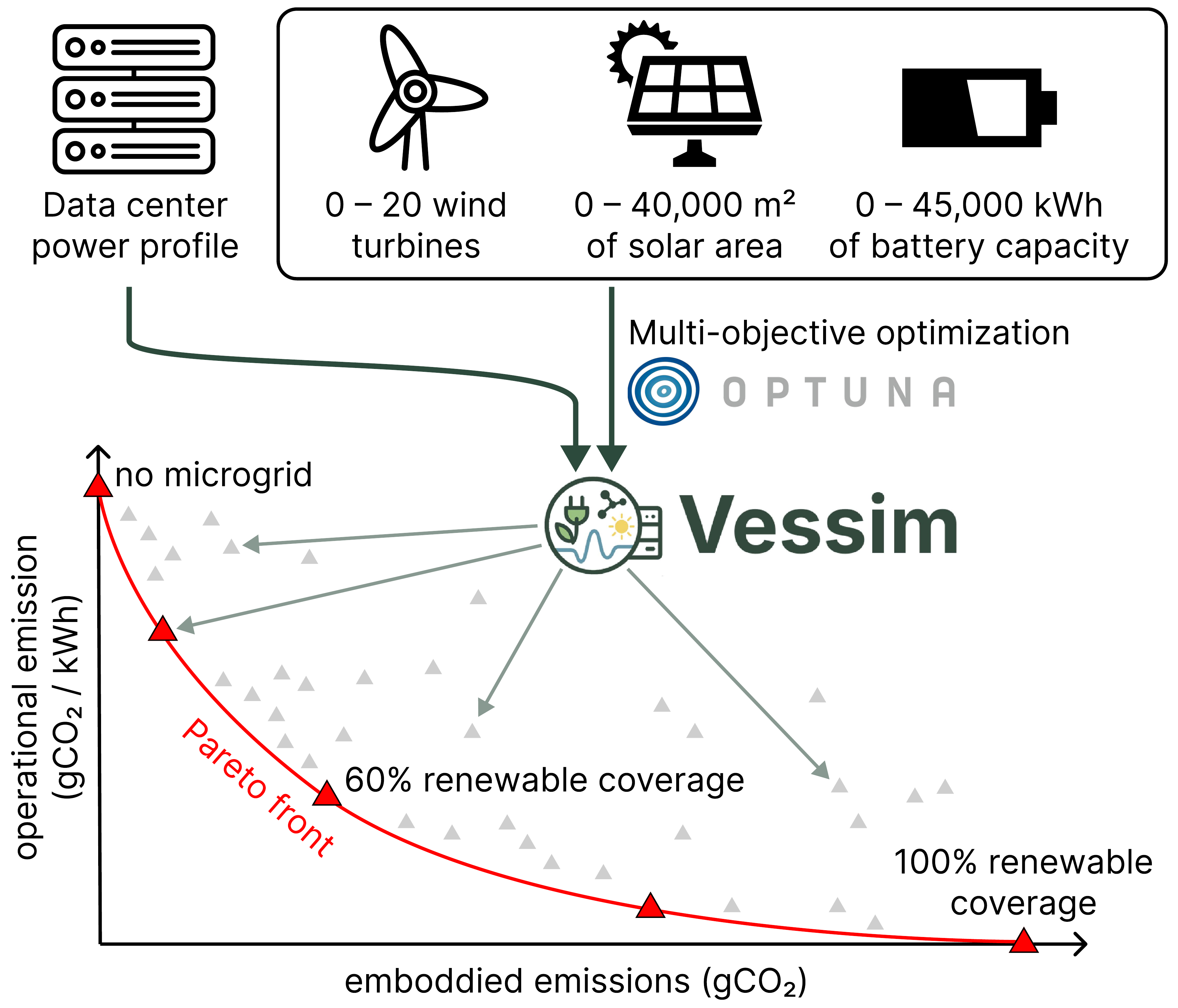}
  \caption{Based on the co-simulator Vessim~\cite{wiesner2024vessim}, we perform a black-box optimization to identify promising microgrid compositions for data centers.}
  \label{fig:fig1}
\end{figure}

On-site energy generation, in particular co-locating data centers with microgrids, offers a promising solution by aligning data center loads with local renewable energy resources, effectively reducing reliance on grid energy. While industry initiatives, like Google's Partnership with Intersect Power and TPG Rise Climate to build renewable-driven data centers~\cite{google2024microgrid}, demonstrate the feasibility of a commitment to co-located renewable energy generation, there is a lack of publicly available tools for modeling and planning such systems.
Existing work has focused either on operational optimization or grid-connected strategies. 
For example, Carbon Explorer~\cite{acun2023} introduces a framework to balance embodied and operational carbon for 24/7 carbon-free data center operation. However, it relies on power purchase agreements and does not support automated right-sizing of on-site renewable microgrids. Other approaches, such as SHIELD~\cite{qi2023}, optimize resource allocation across geo-distributed data centers but do not address the integration or design of on-site energy systems. Similarly, Si et al.~\cite{si2024} propose a blockchain-based control architecture for data center microgrids, focusing on real-time energy dispatch rather than infrastructure optimization.

To address this gap, we present a novel framework for analyzing how different microgrid compositions—specifically the shares of wind power, solar energy, battery storage—affect both the embodied and operational carbon footprint of a specific data center, as illustrated in Figure~\ref{fig:fig1}.
In particular, our contributions are:

\begin{itemize}
    \item We integrate the System Advisor Model (SAM)~\cite{sam2024} into the computing and energy system co-simulator Vessim~\cite{wiesner2024vessim}, enabling high-resolution modeling of solar, wind, and energy storage.
    \item We propose a black-box optimization framework based on the Optuna~\cite{akiba_optuna_2019} framework to support the exploration of large parameter spaces covering microgrid compositions and operational strategies.
    \item We demonstrate our framework using power traces from the Perlmutter supercomputer~\cite{nersc_perlmutter} in two case studies located in Berkeley, CA and Houston, TX, showing how site-specific sizing of renewables and storage influences both operational and embodied carbon emissions, while reducing reliance on grid electricity.
\end{itemize}
All results reported in this paper are fully reproducible. The code and data are available at \url{https://github.com/dos-group/vessim-opt}.

\section{Microgrids for Data Centers}

To improve grid flexibility, microgrids and hierarchical grid structures can help balance supply and demand at a local level. 
A microgrid is a self-contained system that integrates local power generation, energy storage, and demand-side flexibility.
Unlike traditional large-scale grids, microgrids can store and redistribute energy within a small region, reducing dependence on long-distance transmission.
They enable precise demand management through real-time monitoring and control, optimizing energy use and reducing curtailment.

Leading data center operators increasingly adopt microgrids to ensure a stable, low-carbon power supply, as many regional grids struggle with generation capacity limits and transmission bottlenecks~\cite{google2024microgrid, microsoft2024microgrid, migrogridknowledge2024}.
For example, in the US, the time from power project proposal to construction has more than doubled---from under two years for data centers constructed in 2000--2007, to over four years for projects in 2018--2023~\cite{Rand2024QueuedUp}---making grid expansion too slow to keep pace with growing demand~\cite{gorman2025grid_barriers}.
Furthermore, microgrids provide complete control over energy sources.
Google, for example, is investing in renewable-powered microgrids that enable continuous low-carbon operation~\cite{google2024microgrid} without relying on the carbon intensity of the public grid.
Similarly, Microsoft has deployed microgrid-integrated systems that use backup battery power, reducing reliance on fossil-fuel-based generation~\cite{microsoft2024microgrid}.

Recent research on data center microgrids has emphasized optimization under operational uncertainty. For instance, robust multi-objective frameworks have been developed to coordinate wind curtailment, cost, and workload over-provisioning in islanded settings \cite{lian2023}. Similarly, metaheuristic energy scheduling approaches have been used to jointly minimize emissions and electricity costs, but these typically assume a fixed system architecture \cite{khajuria2024}. These works focus primarily on real-time energy management rather than system design and sizing, and rarely address the challenges specific to co-located generation.

Other literature addresses the sizing of hybrid solar-wind-battery systems, using mixed integer optimization or metaheuristics to minimize life cycle cost and improve reliability. Some studies incorporate battery aging constraints \cite{zhao2024}, while others assess the influence of physical parameters such as solar tilt or turbine height on optimal compositions \cite{das2024}. Multi-objective approaches that jointly minimize cost, loss-of-power risk, and carbon emissions have also been proposed for off-grid systems \cite{khemissi2021}. However, these planning models are not tailored to data center workloads and infrastructure.

\section{Exploring Microgrid Compositions via Co-Simulation}
\label{cha:approach}

This section describes the integration of the National Renewable Energy Laboratory's System Advisor Model (SAM)~\cite{sam2024} with the co-simulator Vessim~\cite{wiesner2024vessim} to enable realistic simulations of data center operations within our optimization framework.

\subsection{Microgrid Simulation with Vessim}
Vessim is a recently developed simulation testbed that operates at the interface of computing and energy systems. Built on the Mosaik co-simulation framework~\cite{steinbrink2019}, it enables the composition of heterogeneous simulation models—such as energy producers, consumers, storage units, grid interfaces, and control systems—into integrated microgrid scenarios. Its modular architecture supports both software- and hardware-in-the-loop simulation and allows users to model data center-specific microgrids with fine-grained (e.g. minutely) temporal resolution.

In this study, we simulate the data center’s power demand by feeding historical or forecasted power traces into Vessim’s discrete-event simulation engine. This setup allows us to perform full-year simulations within minutes while preserving the temporal dynamics of power flows and system interactions. To support accurate modeling of on-site renewable integration, we extend Vessim with high-resolution generation models from the SAM, enabling realistic representations of solar, wind, and battery storage systems. This extensibility makes Vessim a suitable foundation for analyzing the behavior and emissions caused by data centers equipped with co-located microgrids.

\subsection{Integrating the System Advisor Model}
The SAM was developed by the National Renewable Energy Laboratory of the U.S. Department of Energy~\cite{sam2024, blair2018} and provides detailed performance and financial modeling capabilities for a wide range of renewable energy systems.
It supports simulations of photovoltaic, wind, geothermal, biomass, marine, solar thermal, industrial process heat, and concentrated solar power systems, as well as hybrid configurations and energy storage.

The SAM provides a software development kit~\cite{samsdk2024} that enables the incorporation of its simulations into software developed in C/C++, C\#, Java, MATLAB, and Python. For the latter, the PySAM package \cite{pysam} is provided, which acts as a wrapper for the SAM Simulation Core (SSC) compute modules. This allows the user to create SAM simulation scenarios directly in their Python applications and without having to run the SAM code generator. The output of the simulation can be directly accessed and evaluated facilitating further analysis or optimization. 

To enable integration of SAM with Vessim, we developed a new component within Vessim's modular signal framework that interfaces directly with SAM's simulation outputs. SAM generates a time series of power production based on a given system configuration, which we map to Vessim’s actor-signal architecture. Specifically, we implemented a dedicated signal class that instantiates and runs a SAM simulation, extracts the resulting power generation profile, and serves time-indexed power values to Vessim actors during simulation. This approach allows SAM-based renewable energy models to be incorporated into co-simulated microgrids within Vessim.

\subsection{Optimization}

Our framework uses Vessim to explore different microgrid compositions across a large parameter space.
In the current implementation, we focus on three key design parameters: the number of wind turbines, the installed solar panel area, and the battery storage capacity. However, the framework is extensible and can incorporate additional technologies such as hydrogen production and storage, and long-duration storage systems like pumped hydro. It can also accommodate different operational strategies such as demand response or carbon-aware scheduling~\cite{wiesner2025qualitytime,Radovanovic_Google_2022,hanafy2024asplos,Souza_Ecovisor_2023}.

\parag{Black-box optimization}
We use Optuna~\cite{akiba_optuna_2019}, a widely adopted framework for black-box optimization, to explore the parameter space of microgrid compositions. Each candidate solution corresponds to a specific combination of wind, solar, and battery resources and is evaluated through a Vessim co-simulation, which models power flows and battery behavior over time to produce detailed operational metrics.
The default optimization objective is to identify a Pareto front that captures the trade-off between operational and embodied carbon emissions. Following the Greenhouse Gas (GHG) Protocol’s Scope~2 definitions, we compute operational emissions as the amount of CO$_2$ released from the purchase of electricity during the facility’s runtime, expressed in tons per day (tCO$_2$/day). In contrast, embodied emissions account for the full lifecycle carbon footprint of infrastructure and are expressed as a one-time investment in tons of CO$_2$ (tCO$_2$). According to the GHG Protocol: “For purposes of accounting for scope 3 emissions, companies should not depreciate, discount, or amortize the emissions from the production of capital goods over time. Instead, companies should account for the total cradle-to-gate emissions of purchased capital goods in the year of acquisition.”~\cite{ghg_scope3_ch3_2013}.

Although our primary focus is on this carbon trade-off, the framework is fully extensible: users may define alternative or additional objectives such as maximizing renewable energy coverage, minimizing battery degradation (e.g., through reduced cycling), or reducing excess energy exports to the grid. These objectives can be optimized individually or in combination, depending on the specific decarbonization goals of the user.

\parag{Extracting candidate solutions}
Optuna explores the parameter space by sampling from user-defined ranges or discrete choices. In the case of multi-objective optimization, it returns a Pareto front of non-dominated solutions representing tradeoffs between competing objectives. 
To aid decision-making, we further process the Pareto front to extract a smaller, representative set of candidate compositions.
This can be achieved through, for example, greedy diversity maximization (which selects solutions that are maximally spread across the solution space), k-means clustering (which groups similar solutions and selects a representative from each cluster), or threshold-based approaches (e.g., returning the best candidates within different embodied carbon budgets). 
By reducing a potentially complex solution space to a small number of diverse and promising options, our framework helps decision-makers to evaluate long-term tradeoffs effectively.

\parag{Implementation}
Our implementation builds on Hydra~\cite{Yadan2019Hydra} in combination with the Optuna sweeper plugin which allows for easy configuration though YAML files and can parallelize the search across a cluster of compute nodes. This setup allows the system to return candidate solutions within a few hours, even for year-long, high-resolution simulations.

\section{Experiments}
\label{cha:evaluation}

We demonstrate our framework on two exemplary scenarios.
For both scenarios, data center power usage is modeled using real power traces from the Perlmutter supercomputer at the National Energy Research Scientific Computing Center (NERSC)~\cite{nersc_perlmutter, zhang2023power}, which exhibits an average power consumption of \SI{1.62}{\mega\watt} during the time period used in our experiments. The simulated data centers are placed in Berkeley, California, and Houston, Texas. These locations were chosen for their contrasting solar and wind resource profiles. The wind data was obtained from the NREL WIND Toolkit \cite{draxl2015wind, draxl2015nrel, king2014validation} and the solar data from the National Solar Radiation Data Base (NSRDB) \cite{buster2025}. %
Our parameter space consists of
\begin{itemize}
    \item A solar farm with rated capacities from \SIrange{0}{40}{\mega\watt} in \SI{4}{\mega\watt} increments, equivalent to an installation area of about 150×150 meters. We assume "low carbon"~\cite{gec2023ulcs} modules of 630\,kgCO$_2$/kW, totaling at 2,520\,tCO$_2$ per increment.
    \item A wind farm with 0 to 10 turbines, each rated at \SI{3}{\mega\watt} with an embodied footprint of 1046\,tCO$_2$ per turbine~\cite{smoucha2016life}.
    \item A battery storage system from 0 to 60\,MWh in 7.5\,MWh increments, which corresponds to one Fluence Smartstack~\cite{fluence2025smartstack}, a state-of-the-art, industry-scale battery unit. Peiseler et al.~\cite{peiseler2024batterycarbon} estimate 62\,kgCO$_2$/kWh for lithium-ion batteries with LFP cathodes, so we assume 465\,tCO$_2$ per battery unit.
\end{itemize}
Solar and wind power generation were modeled using the PVWatts and Windpower modules from the SAM, respectively. 
Battery performance was simulated using the C/L/C battery model by Kazhamiaka et al.~\cite{kazhamiaka2019}, which is already integrated in Vessim.
We performed an exhaustive search over the defined parameter space to establish a baseline for subsequent analyses.

\subsection{Exploring the Trade-Off Between Operational and Embodied Emissions}

Planning microgrids alongside on-site power supply can significantly reduce operational emissions.
However, it also leads to additional embodied emissions related to the production, maintenance, and disposal of renewable energy systems and storage.
In this chapter, we provide an exemplary analysis of operational and embodied carbon emissions for both data center locations.
Since our focus is on a carbon accounting use case, we calculate operational carbon emissions using average carbon intensity data from Electricity Maps~\cite{electricitymaps_caiso_2025, electricitymaps_erco_2025}, rather than alternative metrics such as marginal carbon intensity~\cite{wiesner2025moving}.

\begin{figure}
  \centering
  \includegraphics[width=\columnwidth]{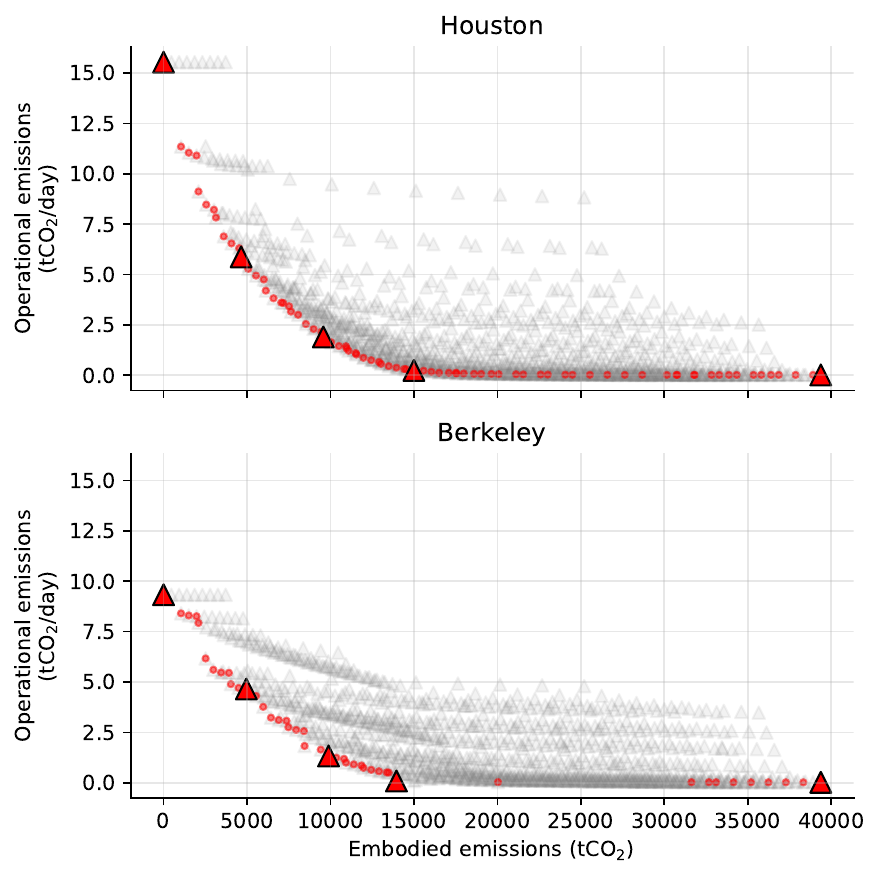}
  \vspace{-6mm}
  \caption{Pareto-front (red dots) and candidate solutions (red triangles) when optimizing for embodied and operational carbon emissions simultaneously.}
  \label{fig:pareto}
\end{figure}

Figure~\ref{fig:pareto} illustrates the Pareto fronts for both scenarios. 
Each point on the front represents a composition that cannot improve on one dimension (e.g., operational emissions) without incurring a penalty in the other (e.g., embodied emissions).
For both locations, we observe that moving toward close-to-zero operational emissions requires a substantial increase in embodied emissions, which is in line with findings from related works~\cite{acun2023}.

To facilitate a more interpretable comparison of design strategies, we selected a set of five representative compositions per location, shown in Tables~\ref{tab:houston} and \ref{tab:berkeley}. These include a baseline composition without any on-site generation or storage, the compositions with lowest operational emissions under embodied emission constraints of up to 5,000, 10,000, and 15,000\,tCO$_2$, plus the composition with the lowest operational emissions overall. 

The shape of the Pareto fronts reflects the renewable resource profiles of each site.
In Houston, the first candidate represents a fully grid-powered data center with no renewable infrastructure and consequently the highest operational emissions at 15.54\,tCO$_2$/day. The next three rows show steady progress in operational decarbonization with increasing investment. For example, the best composition below 5,000\,tCO$_2$ (12\,MW wind, 7.5\,MWh battery) cuts operational emissions by more than half, while the 14,999\,tCO$_2$ composition (12\,MW wind, 12\,MW solar, 52.5\,MWh battery) achieves more than 99\,\% on-site coverage. The lowest operational emissions are achieved by the most carbon-intensive composition, combining maximum wind and solar capacity with full storage, at a steep embodied cost of 39,380\,tCO$_2$. These results underscore the challenge of decarbonizing in regions where variable and less favorable wind conditions require more aggressive infrastructure deployment to approach net-zero operation.

In contrast, Berkeley benefits from stronger and more consistent renewable resources, especially solar, allowing it to reach near-zero operational emissions at lower embodied cost. The best composition under 5,000\,tCO$_2$ (3\,MW wind, 4\,MW solar, 22.5\,MWh battery) already reduces emissions by over 50\,\% relative to the baseline. At 9,885\,tCO$_2$, a fully solar-powered system with adequate storage (12\,MW solar, 37.5\,MWh battery) enables almost 92\,\% coverage, and with 13,953\,tCO$_2$ investment, the system covers 99.57\,\% of its load. The final, unconstrained composition mirrors Houston’s in component sizing but achieves low operational emissions (0.02\,tCO$_2$/day) with significantly higher efficiency in terms of coverage and battery utilization. 
This shows Berkeley’s suitability for early, low-cost carbon reductions and demonstrates how renewable performance characteristics should influence infrastructure decisions.

\begin{table}
    \small
    \caption{Houston candidate solutions.}
    \vspace{-4mm}
    \begin{tabular}{rrrrrrr}
    \toprule
    \multicolumn{3}{c}{Composition} & \multicolumn{4}{c}{Resulting metrics} \\
    \cmidrule(lr){1-3} \cmidrule(lr){4-7}
    Wind & Solar & Battery & Embodied & Operat. & Cov. & Battery \\
    (MW) & (MW) & (MWh) & (tCO$_2$) & (tCO$_2$/d) & (\%) & cycles \\
    \midrule
    0 & 0 & 0.0 & 0 & 15.54 & 0.00 & -- \\
    12 & 0 & 7.5 & 4,649 & 5.88 & 71.07 & 153 \\
    9 & 8 & 22.5 & 9,573 & 1.90 & 91.79 & 129 \\
    12 & 12 & 52.5 & 14,999 & 0.24 & 99.11 & 71 \\
    30 & 40 & 60.0 & 39,380 & 0.02 & 100.00 & 41 \\
    \bottomrule
    \end{tabular}
    \label{tab:houston}
    \vspace{2mm}

    \caption{Berkeley candidate solutions.}
    \vspace{-4mm}
    \begin{tabular}{rrrrrrr}
    \toprule
    \multicolumn{3}{c}{Composition} & \multicolumn{4}{c}{Resulting metrics} \\
    \cmidrule(lr){1-3} \cmidrule(lr){4-7}
    Wind & Solar & Battery & Embodied & Operat. & Cov. & Battery \\
    (MW) & (MW) & (MWh) & (tCO$_2$) & (tCO$_2$/d) & (\%) & cycles \\
    \midrule
    0 & 0 & 0.0 & 0 & 9.33 & 0.00 & -- \\
    3 & 4 & 22.5 & 4,961 & 4.65 & 60.11 & 82 \\
    0 & 12 & 37.5 & 9,885 & 1.33 & 91.85 & 206 \\
    9 & 12 & 52.5 & 13,953 & 0.08 & 99.57 & 138 \\
    30 & 40 & 60.0 & 39,380 & 0.02 & 99.95 & 106 \\
    \bottomrule
    \end{tabular}
    \label{tab:berkeley}
\end{table}

\subsection{Projecting Long-Term Emissions}
\label{sec:longterm_projection}

\begin{figure}
  \centering
  \includegraphics[width=\columnwidth]{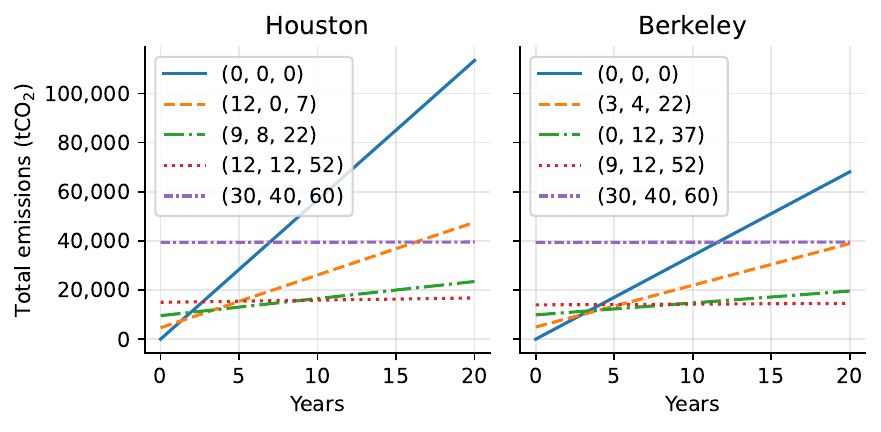}
  \vspace{-7mm}
  \caption{Naive 20 year projection of total data center emissions for the five candidate solutions, denoted by (wind capacity in MW, solar capacity in MW, battery capacity in MWh).}
  \label{fig:longterm}
\end{figure}

To assess the long-term impact of on-site renewable investments, we project cumulative emissions over a 20-year horizon based on each composition’s embodied and operational carbon profile. Figure~\ref{fig:longterm} shows this projection for the five representative solutions per site (see Tables~\ref{tab:houston} and \ref{tab:berkeley}). Each line begins at the respective composition’s embodied emissions and accumulates operational emissions over time, assuming a constant daily emissions rate and no reinvestments.

Our results demonstrate that minimizing operational emissions at all costs is not necessarily optimal over the system's lifetime. In both regions, the “zero operational carbon” configurations (those with the largest on-site deployments) start with such high embodied emissions of 39,380\,tCO$_2$ that they remain among the most carbon-intensive solutions, even after 20 years of operation.
Conversely, the “zero embodied carbon” baseline (0\,MW wind, 0\,MW solar, 0\,MWh battery) accumulates emissions quickly, becoming the worst-performing configuration after approximately 7 years in Houston and 12 years in Berkeley.

These trajectories underscore that the optimal level of renewable deployment depends not only on local resource availability but also on the expected operational lifetime of the facility. For short-lived systems, minimal or moderate deployment may lead to lower total emissions. In contrast, for deployments expected to run over decades, more ambitious investments can be justified.

It is important to note that this analysis assumes all components have equal lifetimes. 
In practice, solar panels and wind turbines often last over 20 years, whereas batteries may require replacement within 10--15 years.
Since we do not model reinvestment or degradation, the analysis represents a conservative baseline. Still, it highlights a central planning challenge: \emph{achieving 100\,\% on-site coverage is not inherently optimal}. 
Effective planning must balance ambition with expected system lifetime and regional conditions.

\subsection{Optimization Beyond Carbon Emissions}

\begin{figure}
  \centering
  \vspace{-5mm}
  \includegraphics[width=.8\columnwidth]{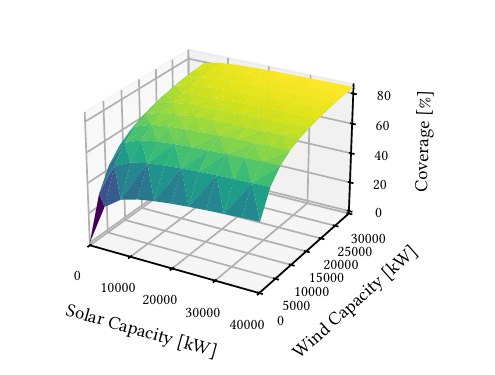}
  \vspace{-4mm}
  \caption{On-site renewable coverage for different compositions (without batteries) in Houston.}
  \label{fig:houston_3d}
\end{figure}

The flexibility of our framework allows users to optimize for a wide range of additional objectives. Figure~\ref{fig:houston_3d} illustrates one such dimension: On-site renewable coverage, i.e., the fraction of demand directly met by on-site solar and wind generation.
This specific analysis excludes battery storage to isolate the impact of generation capacity. In Houston, renewable coverage improves with increasing wind and solar capacities, though with diminishing returns at higher deployment levels. Such visualizations help identify sweet spots in the solution space where relatively small investments yield significant improvements.

Other metrics and possible optimization objectives include:
\begin{itemize}
    \item Battery degradation minimization, where the goal is to reduce wear and prolong battery lifespan, e.g., by avoiding frequent deep cycling~\cite{BLASTLite, o2022lithium}.
    \item Electricity cost reduction, especially in regions with volatile grid pricing or time-of-use tariffs, where shifting or shaving demand can yield economic savings.
    \item Load shifting potential, to evaluate how well the system can adapt to flexible or deferrable demand by implementing carbon-aware scheduling policies in Vessim.
    \item Reliability or resilience metrics, particularly relevant for off-grid or backup scenarios, measuring the fraction of time the system can operate independently of the grid.
\end{itemize}
In future work, more advanced objective functions and surrogate-assisted search strategies could be integrated to explore complex trade-offs across multi-dimensional objectives in large-scale, realistic deployments.

\subsection{Search Performance}

To assess the efficiency of our optimization framework, we compare it against an exhaustive search over the full parameter space. The baseline evaluates all 1089 valid combinations of solar, wind, and battery compositions, each requiring a full-year, high-resolution co-simulation—resulting in runtimes exceeding 24 hours, even on parallel infrastructure.
In contrast, our black-box approach uses only 350 trials with a population size of 50, guided by the NSGA-II algorithm~\cite{deb2002}. Despite this drastic reduction in search effort, it successfully recovers around 80\,\% of all Pareto-optimal solutions, demonstrating a $\sim$2.4× speed-up. The process is fully parallelizable through Hydra and Optuna and scales easily across multi-core systems or clusters.

This remains a preliminary analysis based on illustrative scenarios and a limited parameter space. Future work should evaluate more advanced search strategies, such as dynamic pruning or early stopping for non-proimising simulation runs, applied to larger, more realistic deployments with broader parameter spaces.

\section{Conclusion}
\label{cha:conclusion}

In this paper, we presented a novel optimization framework designed to support the planning of colocated data center microgrids.
By integrating high-resolution renewable energy models from the System Advisor Model (SAM) into the Vessim co-simulation platform and coupling it with a black-box optimization backend, our approach enables automated sizing of solar, wind, and battery systems to balance performance and carbon impact.

Our evaluation across two different geographic scenarios highlights that optimal microgrid design is inherently location-specific. %
By offering a flexible and extensible foundation, our framework sets the stage for more informed and systematic decision-making in the design of sustainable energy systems for data centers.

\begin{acks}
Funded by the Deutsche Forschungsgemeinschaft (DFG, German Research Foundation) – Project-ID 414984028 – SFB 1404 FONDA.
\end{acks}

\bibliographystyle{ACM-Reference-Format}
\bibliography{bibliography}

\end{document}